\documentclass[preprint,showpacs,preprintnumbers,amsmath,amssymb]{revtex4}
\usepackage{graphicx}
\usepackage{dcolumn}
\usepackage{bm}
\begin{document}

\title{ Thermodynamics of a quantized electromagnetic field
in rectangular cavities with perfectly conducting walls}

\author{R. J\'{a}uregui, C. Villarreal and S. Hacyan}
\affiliation{Instituto de F\'{\i}sica, Universidad Nacional
Aut\'onoma de M\'exico, Apartado Postal 20-364, M\'exico, 01000,
D.F., Mexico}
\date{\today}
\begin{abstract}
The thermodynamical properties of a quantized electromagnetic
field inside a box with perfectly conducting walls are studied
using a regularization scheme that permits to obtain finite
expressions for the thermodynamic potentials. The source of
ultraviolet divergences is directly isolated in the expression for
the density of modes, and the logarithmic infrared divergences are
regularized imposing the uniqueness of vacuum and, consequently,
the vanishing of the entropy in the limit of zero temperature. We
thus obtain corrections to the Casimir energy and pressures, and
to the specific heat that are due to temperature effects; these
results suggest effects that could be tested experimentally.
\end{abstract}
 \maketitle

\section{Introduction}
The theory of blackbody radiation in cavities has played a
decisive role in the development of quantum physics. In
particular, the existence of a zero-point energy \cite{milloni}
with measurable physical effects, as predicted by Casimir in 1948
\cite{casimir}, followed from its consistency. It is now well
established that the fluctuations of the quantum vacuum induce an
attractive force of magnitude (per unit area) $\pi^2 \hbar c/240
L^4$ between two perfectly-conducting parallel plates separated a
distance $L$. The physical importance of the Casimir effect
prompted an extensive investigation of the fundamental properties
of quantum and thermal fluctuations for a large variety of systems
with particular dielectric and geometric features\cite{early,
mehra,brown}. Nevertheless, a complete self-consistent theory of
quantum fluctuations in closed cavities is far from being achieved
due to the lack of a clear and well established regularization
scheme.

In the simplest case of an ideal conducting parallel-plate
configuration at finite temperature, relevant thermodynamical
quantities such as entropy, internal and free energies, and
pressure forces have been calculated by many authors following
different lines \cite{mehra,brown,ambjorn,revzen,genet}. In a
pioneering work based on general symmetry considerations, Brown
and Maclay \cite{brown} analyzed the structure of the
electromagnetic stress-energy tensor $T^{\mu\nu}$ for two parallel
plates immersed in a thermal bath at a temperature $T$. They found
that this tensor can be written as a sum of three terms:
\begin{equation}\label{maclay1}
 T^{\mu \nu}= T^{\mu \nu}_{0;L} +  T^{\mu \nu}_{T;\infty}
+ T^{\mu \nu}_{T;L},
\end{equation}
where the first term represents the zero-point Casimir stress
tensor for a plate separation $L$, the second is due to the black
body radiation between widely separated plates, and the third term
is a correction that vanishes in the limits $T \rightarrow 0$ and
$L\rightarrow \infty$. The resulting energy density coincides with
the standard Stefan-Boltzmann expression, $E/V= T^{0 0}_{T;\infty}
=(\pi^2/15 \hbar^3 c^3) (k_BT)^4$, but there is also a
contribution due to boundary effects in the limit $T\rightarrow
\infty$: the pressure at the plates deviates from the standard
blackbody pressure by a term that increases linearly with the
temperature. This additional term is independent of $\hbar$
\cite{brown}; in fact, Boyer \cite{boyer} also derived it within a
purely classical formalism, and Mehra \cite{mehra}, and Feinberg
$et$ $al.$ \cite{revzen} obtained identical results. This linear
term has also been found for other geometries and materials
\cite{hoye, jaffe}.

There are  some  controversial issues regarding the Casimir forces
between dielectric parallel plates. For instance, the Lifshitz
formula \cite{lifshitz} that follows from a direct application of
Matsubara's frequencies formula is ambiguous in the infrared
regime. The fundamental issue seems to be whether the transverse
electric (TE) component of the electromagnetic field contributes
to the Casimir force in that regime. The theoretical answer
depends on the model used to characterize the dielectric response
of the material: at low frequencies, the TE mode vanishes if the
Drude model for the dielectric function is used \cite{bostrom},
but this is not the case of the plasma model \cite{bordag}.
According to the first approach, there must be a large thermal
correction to the Casimir force between plates with a separation
of one micrometer or larger, but no such effect has been observed
in experiments \cite{lamoreaux, torgerson}. Moreover, Bezerra $et$
$al.$ \cite{bezerra} have shown that using the Drude model, the
entropy turns out to be finite at zero temperature and depends on
the parameters of the system, thus violating the third law of
thermodynamics; these authors proposed to solve the problem with
the plasma model and the Leontovich surface impedance approach,
but the applicability of this scheme has been questioned by others
\cite{esquivel, svetovoy4,milton}. Still other authors
\cite{torgerson,milton,svetovoy} have concluded that some of the
discrepancies are due to the fact that, in the low frequency
regime, the atomic nature of the material invalidates the use of
dielectric functions that characterize bulk properties.

The problem becomes considerably more complicated beyond the
parallel plates configuration. In their pioneering work, Case and
Chiu \cite{case} used the fluctuation-dissipation theorem to
calculate the mean internal energy inside a perfectly conducting
cavity filled with a dielectric dispersive medium at finite
temperature. For an empty cubic cavity, they obtained an
expression for the internal energy per unit volume similar to
Eq.(\ref{maclay1}), and they also found a deviation from the
Stefan-Boltzmann law at low temperatures. However, Case and Chiu
did not calculate other thermodynamic potentials that also
characterize the system.

In a later work, Ambjorn and Wolfram \cite{ambjorn} studied the
general problem of the constrained vacuum fluctuations of $scalar$
fields in D-dimensional Euclidean spaces, subject to either
Dirichlet or Neumann boundary conditions along $P$ orthogonal
directions; they calculated the partition function of such a
configuration with a dimensional regularization technique.
However, the resulting partition function for a completely closed
cavity ($D=P$) turned out to be divergent, as well as
thermodynamic quantities such as the free energy and the entropy.
Ambjorn and Wolfram claimed that these divergences could be
removed introducing two further regularization schemes, valid at
low and high temperatures separately, but they gave no explicit
expressions for the finite thermodynamic potentials that may be
valid at $all$ temperatures. A similar problem also arises when
dealing with $electromagnetic$ fluctuations in rectangular closed
cavities, as shown by Santos and Tort \cite{santos} who studied
the asymptotic behavior of the free energy at low and high
temperatures using a duality inversion formula. Such a duality
symmetry was formerly discovered by Brown and Maclay for the
parallel plate configuration \cite{brown} and studied by Ravndal
and Tollefsen \cite{ravndal}. More recent works along these lines
are due to Inui \cite{inui} and Cheng \cite{cheng}. However, none
of these authors report the free energy of the EM field in a
closed and finite form that could be valid for all temperatures
and box geometries. As a consequence, no quantitative predictions
for temperature corrections to the Casimir pressures can be
deduced from these works.

The aim of the present paper is to study the thermal effects of a
quantum electromagnetic field in closed rectangular cavities with
perfectly conducting walls. For this purpose, we propose a
regularization scheme that permits to calculate the
thermodynamical variables and obtain finite results that can be
eventually tested in laboratories. The analogous problem in the
zero-temperature case has been addressed in several previous works
\cite{ambjorn,lukosz,hacyan93,maclay}, the main results being that
the Casimir energy and pressures on the cavity plates depend
strongly on the geometry (they can even change sign according to
the relative ratios of the faces). In general, the Casimir
pressures produce instabilities and, in the particular case of a
cube, the forces acting on its faces are repulsive\cite{hacyan93}.
In the present paper we generalize the previous results to include
a thermal bath in the configuration. In Section 2, the
regularization is achieved using a particular representation of
the EM mode density that permits to isolate the contributions of
ultraviolet divergences (arising from zero-point fluctuations, to
the internal and free energies). For this purpose, we first
isolate the contribution to the mode density in a large cavity, as
given by Weyl's asymptotic expression, including correction terms
proportional to the edges of the configuration. Next, the infrared
divergences are regularized by imposing the uniqueness of the
vacuum state, $i.e.$ the state without photons of any frequency,
in such a way that the entropy of the system at zero temperature
is zero by construction. The regularized free energy is then used
to calculate the pressures, which are physical observable
quantities in experiments. In Section. 3, our results are compared
with other approaches in order to elucidate the origin of the
difficulties arising in the problem of quantum and thermal
fluctuations in rectangular cavities.

\section{Thermodynamic variables}
For any given field at thermal equilibrium with its surroundings,
the contribution to the free energy of a mode of frequency
$\omega_{\bf k}$ is
\begin{equation}
F(\omega_{\bf k}) = T\ln(1-e^{-\hbar\omega_{\bf k}/T}) +
\frac{1}{2}\hbar\omega_{\bf k} ~,\label{1}
\end{equation}
in standard notation, where the second term is the contribution of
the zero-point field (from now on, we set $k_B=1$). The total free
energy is a sum over the set of modes:
\begin{equation}
F= \int \rho(\omega_{\bf k}) F(\omega_{\bf k}) d\omega_{\bf k}~,
\end{equation}
where $\rho(\omega_{\bf k})$ is the density of modes.

If we consider a rectangular cavity with sizes $a_k$ ($k=1,2,3$)
made of an idealized perfect conducting material, the boundary
conditions imply the discretization of the frequencies in the box:
\begin{equation}
\omega_{\bf n} = \pi c \sqrt{\Big(\frac{ n_1}{a_1}\Big)^2
+\Big(\frac{n_2}{a_2}\Big)^2+\Big(\frac{n_3}{a_3}\Big)^2}
\label{omega}.
\end{equation}
The density of modes has the form \cite{hacyan93,case,santos}
\begin{equation}
\rho(\omega) = \frac{1}{4} {\sum_{\bf n}}^\prime \delta (\omega -
\omega_{\bf n})( 1 - \delta_{n_10}\delta_{n_20}-
\delta_{n_20}\delta_{n_30} -
\delta_{n_30}\delta_{n_10})\label{eq:rho}~,
\end{equation}
where the prime in the summation indicates that the term with all
three indices $n_i=0$ is to be excluded. This particular form of
$\rho(\omega)$ reflects the fact that the normal (parallel)
components of the magnetic (electric) field vanish on the walls of
the cavity, and that the electromagnetic field admits two
polarization states represented by transverse electric (TE) and
transverse magnetic (TM) modes. These modes may be derived from
massless scalar Hertz potentials that satisfy mixed combinations
of Dirichlet or Neumann boundary conditions on the cavity plates
\cite{hacyan93,dalvit}.

As shown in Ref.~\cite{hacyan93}, ultraviolet divergences of the
zero-point energy can be isolated. For this purpose, the spectral
density is written in terms of a summation involving the variable
\begin{equation}
u_{\bf n} = \frac{2}{c}\Big[(n_1 a_1)^2 + (n_2 a_2)^2 + (n_3
a_3)^2\Big]^{1/2}~,
\end{equation}
which is conjugate to the frequency  variable $\omega$. As shown
in Ref.~\cite{hacyan93}, it is convenient to separate the density
of modes into two parts:
\begin{equation} \rho(\omega)
=\rho^{\infty}(\omega)+ \Delta \rho(\omega) ,
\end{equation}
where the first term is
\begin{equation}
\rho^{\infty}(\omega) =  \frac{V}{\pi^2c^3}\omega^2 -\frac{1}{2\pi
c} (a_1 + a_2 + a_3)~,
\end{equation}
and the second term can be written in the form
\begin{eqnarray} \Delta
\rho(\omega) &=& \frac{V}{\pi^2c^3} {\sum_{{\bf n}
=-\infty}^\infty} ^\prime \frac{\omega\sin(\omega ~u_{\bf
n})}{u_{\bf n}} - \frac{a_1}{2\pi c} {\sum_{n=-\infty}^\infty}
^\prime\cos (\omega~ u_{n00})\nonumber
\\
&-& \frac{a_2}{2\pi c} {\sum_{n=-\infty}^\infty} ^\prime\cos
(\omega~ u_{0n0})- \frac{a_3}{2\pi c}
{\sum_{n=-\infty}^\infty}^\prime \cos (\omega~ u_{00n}) ~.
\label{eq:rhop}
\end{eqnarray}
The above relations are valid for all values of $\omega \ne 0$.

The variable $c u_{\bf n}$ in the above formula can be interpreted
as the position of the image charges that generate the field
correlation functions with the appropriate boundary conditions on
the conducting walls \cite{hacyan93,hacyan90}. Moreover, since
these variables are conjugate to the frequency, lower values of
the integer $n_i$'s in $u_{\bf n}$ correspond to higher values of
the frequency, and vice versa. Notice also that Eq.
(\ref{eq:rhop}) is not well defined for $\omega = 0$ because the
terms related to edges diverge, while in the original expression
for $\rho(\omega)$, Eq.~(\ref{eq:rho}), the term with $\omega=0$
is absent.

The spectral density $\rho^{\infty}(\omega)$ prevails in the limit
$a_i\rightarrow \infty$, but it is not identical to the density of
modes in free space because the edge terms are important even in
this limit. However, just for simplicity, we shall call it ``free"
or ``blackbody" density . The use of $\Delta \rho(\omega)$ instead
of the full density of states $\rho(\omega)$ for evaluating an
extensive variable is equivalent to calculating the ${\it
difference}$ between the values of that variable in bounded and in
``free" space.

The contribution of the zero-point field to either the free or the
internal energy associated to $\rho^\infty(\omega)$ is always
divergent. However, the temperature dependent part of the free
energy that corresponds to the blackbody radiation is perfectly
finite and is given by
\begin{eqnarray}
F^\infty_{BB} &\equiv& T\int_0^\infty \rho^\infty(\omega)\log (1 -
e^{-\hbar\omega/T})d\omega\nonumber\\
 &=&-\frac{\pi^2 }{45\hbar^3c^3}V T^4
-\frac{\pi^2}{12\hbar c}(a_1+a_2+a_3)T^2~ . \label{eq:FBB}
\end{eqnarray}
The first term, proportional to the volume of the box, is the
usual text-book formula; the second term is the contribution of
the cavity edges to the free energy \cite{pathria}. It is easy to
see from a dimensional analysis that any contribution to the free
energy proportional to the surface of the walls $a_ia_j$ should be
proportional to $T^3$, and similarly a contribution related to the
vertices should be proportional to $T$ and independent of the
parameters $a_i$. As mentioned above, contributions to the
spectral density from faces and vertices of the cavity do not
appear due to a cancellation of the corresponding TE and TM modes
\cite{hacyan93,pathria}. In any case, a correction of the Stefan-
Boltzmann law, possibly as the one given by Eq.~(\ref{eq:FBB}),
has been observed during the development of masers
technology\cite{siegman}.

The contribution of the spectral density $\Delta \rho(\omega)$ to
the zero-point free or internal energy is the Casimir term
\begin{eqnarray}
\Delta E_0 &=&\frac{\hbar}{2}\int_0^\infty \omega\Delta\rho(\omega)d\omega \nonumber \\
&=&-\frac{\hbar}{\pi^2c^3}V
{\sum_{\bf n}}^\prime\frac{1}{u_{\bf n}^4}\nonumber\\
&+&\frac{\hbar}{4\pi c}a_1{\sum_n}^\prime\frac{1}{u_{n00}^2}
+\frac{\hbar}{4\pi c}a_2{\sum_n}^\prime\frac{1}{u_{0n0}^2}
+\frac{\hbar}{4\pi c}a_3{\sum_n}^\prime\frac{1}{u_{00n}^2}~.
\end{eqnarray}
The thermal part of the free energy calculated with $\Delta \rho
(\omega)$ turns out to be free of ultraviolet divergences.
However, some of the integrals involved do not have a unique
interpretation as shown in the Appendix. In fact,
\begin{eqnarray}
\Delta F &=& \Delta E_0 +  \frac{\pi^2}{2 \hbar^3c^3}VT^4
{\sum_{\bf n}}^\prime h_V(v_{\bf n})\nonumber\\ &+& \frac{\pi}{4
\hbar c} T^2 {\sum_n}^\prime \big[a_1 f_E(v_{n00}) + a_2
f_E(v_{0n0}) + a_3 f_E(v_{00n})\big]~, \label{eq:deltaF}
\end{eqnarray}
where we have defined
\begin{eqnarray}
f_{V,E}(v) &=& \frac{1}{v}\Big(g(v) - K_{V,E}(v)\Big)~, \nonumber \\
h_{V,E}(v) &=& h(v) + \frac{1}{v^3}K_{V,E}(v)~,\nonumber\\
g(v) &=& \coth(v) -\frac{1}{v}~,\nonumber\\
h(v) &=&\frac{1}{v}\frac{d}{dv}\Big(\frac{1}{v}g(v)\Big)
\end{eqnarray}
as functions of the dimensionless variable
$$
v_{\bf n}=\frac{\pi T}{\hbar} u_{\bf n}
$$
(at room temperature $k_B T \sim 2.6 \times 10^{-2}$ eV, and this
implies that $ \hbar c /\pi k_B T \sim 3 \mu$m ). $K_V(v)$ and
$K_E(v)$ are trichotomic functions of $v$ that can take only the
values $\pm 1$ and $0$, depending on the contour of integration
selected to evaluate the integrals appearing in the volume and
edge parts, respectively, of the free energy $\Delta F$. Notice
that in the limit $T\rightarrow 0$, the free energy $\Delta F$ is
equal to the internal energy $\Delta E_0$ independently of the
value of $K$.

 In general, given an integral expression for a physical variable, the
selection of an integration path corresponds to a particular
boundary condition for that variable. In the following, we shall
impose the limiting condition
\begin{equation}
\frac{\partial F}{\partial T}{\Big \vert}_{T\rightarrow 0^+} = 0~,
\label{eq:unique}
\end{equation}
in order to select an appropriate integration path for a given
value of $v_{\bf n}$.
 The entropy ${\cal S}$ of the system being
\begin{equation}
{\cal S} \equiv -\frac{\partial F}{\partial T} = {\cal
S}^\infty_{BB}+ \Delta {\cal S}~,
\end{equation}
Eq. (\ref{eq:unique}) implies the uniqueness of the EM vacuum. The
blackbody contribution  to the entropy is
\begin{equation}
{\cal S}^\infty_{BB} = \frac{4 \pi^2}{45\hbar^3c^3}V T^3
+\frac{\pi^2}{6\hbar c}(a_1+a_2+a_3)T~,
\end{equation}
and using the integrals given in the Appendix, the remaining part
of the entropy can be written as
\begin{eqnarray}
\Delta{\cal S} &=& - \frac{\pi^2 VT^3}{2\hbar^3 c^3}
{\sum_{n_i}}^\prime\Big[ v_{\bf n}h^\prime(v_{\bf n})
-\frac{3}{v_{\bf n}^3}K_V(v_{\bf n}) +4 h_V(v_{\bf n})
\Big]\nonumber\\
&-&\frac{\pi a_1 T}{4\hbar c}{\sum_n}^\prime
\Big[2f_E(v_{n00}) + v^2_{n00}h_E(v_{n00})\Big]\nonumber \\
&-&\frac{\pi a_2T}{4\hbar c}{\sum_n}^\prime
\Big[2f_E(v_{0n0}) + v^2_{0n0} h_E(v_{0n0})\Big]\nonumber \\
&-&\frac{\pi a_3T}{4\hbar c}{\sum_n}^\prime \Big[2f_E(v_{00n}) +
v^2_{00n}h_E(v_{00n})\Big].
\end{eqnarray}

In order to evaluate the limit $T\rightarrow 0$ of the above
formulas, the summation over discrete indices can be replaced by
an integration over a continuous dummy variable $v$:
\begin{equation}
{\cal S}\rightarrow - \frac{1}{4}\int_0^\infty dv \Big[v^2 [v
h'(v) -\frac{3}{v^3}K_V(v) + 4 h_V(v)]+6 f_E(v) +3 h_E(v)\Big].
\end{equation}
This expression is independent of the geometric parameters $a_i$
as expected from Nernst's third law of thermodynamics. The direct
evaluation of these integrals gives the result
\begin{eqnarray}
{\cal S} &&\rightarrow - \frac{1}{4} \int_0^\infty
dv\frac{1}{v}\Big(\coth(v) - K_V(v) -\frac{1}{v}\Big) \nonumber \\
&&- \frac{3}{4}\Big[1 + \int_0^\infty dv\frac{1}{v}\Big(\coth(v) -
K_E(v) - \frac{1}{v}\Big)\Big] , \label{eq:Seq0}
\end{eqnarray}
where the first term comes from the contribution of the volume
terms, and the second from the edge terms. Now, if $K(v)$ is taken
strictly either as $0$, or as $1$ or as $-1$ for any value of $v$,
the entropy at zero temperature will not be zero; even worse, it
will clearly diverge. If $K=0$, the integrands would be regular at
$v =0$, while the edge and volume integrals would have a
logarithmic divergence. On the other hand, if $K$ equals 1 or -1,
the integrand would not be regular at $v=0$ and the integrals
would diverge. Thus, the simplest way of keeping a bounded value
for the entropy is to choose $K=0$ for $v <v_0$ and $K=1$ for
larger values of $v$, with $v_0$ to be determined. Accordingly, we
define the function
\begin{equation}
G(v_0) =\int_0^{v_0} dv\frac{1}{v}\Big(\coth(v)
-\frac{1}{v}\Big)+\int_{v_0}^\infty dv\frac{1}{v}\Big(\coth(v)
-\frac{1}{v} - 1 \Big),
\end{equation}
and choose a value of $v_0$ in such a way that the entropy does
take the value zero at $T=0$. A plot of $G(v_0)$ is shown in
Fig.~1; it turns out to be a increasing function of $v_0$, such
that $G(v_V)=0$ and $G(v_E)=-1$ for $v_V=1.763876988$ and
$v_E=0.64889408$ (with ten significant figures).

Thus choosing $K_{V,E}=\Theta (v- v_{V,E})$, where $\Theta$ is the
standard step function, we guarantee that $\Delta F$ satisfy the
boundary condition Eq.~(\ref{eq:unique}) and that a finite
regularized value of the thermodynamic variables $\Delta F$ and
$\Delta S$ is obtained.

At this point, we recall that the spectral density given by
Eq.~(\ref{eq:rhop}) is not valid for $\omega=0$; however, if this
equation is taken as it stands for all values of $\omega$, it
includes a term proportional to $\delta (\omega)$. It is precisely
such a term that gives rise to the extra factor $3/4$ appearing in
the contributions of the edges to the entropy at zero temperature
in Eq.~(\ref{eq:Seq0}). This unphysical contribution to the
entropy is properly removed by the selection of $v_E$.

From a physical point of view, it is not possible in general to
distinguish between processes that differ from each other by the
emission of a certain number of low energy photons with
$\omega\rightarrow 0$. In free space, this is true for a
continuum, but in a closed box, the values of $v_{V,E}$ are the
ones that define which photons should be taken as being of low
energy; namely,
\begin{equation}
v_{\bf n}> v_V \Leftrightarrow k_B T
>\frac{\hbar c v_V}{u_{\bf n} \pi}, \label{desig}
\end{equation}
and this condition determines whether the photons contribute to
the counting of microstates. That is to say, the uniqueness of the
vacuum determines which modes are to be considered as infrared.

Given $S$ and $F$, the expression for the internal energy can be
directly calculated as
\begin{equation}
E(T) = E_{BB}(T) + \Delta E(T)~,\nonumber \\
\end{equation}
where
\begin{equation}
E^\infty_{BB}=\frac{\pi^2 }{15\hbar^3c^3}V T^4
+\frac{\pi^2}{12\hbar c}(a_1+a_2+a_3)T^2\nonumber\\
\end{equation}
and
\begin{eqnarray}
\Delta E(T) &=& \Delta F + T\Delta {\cal S}\nonumber \\
&=& \Delta E_0 - \frac{\pi^2}{2 \hbar^3 c^3}VT^4
{\sum_{a_1a_2a_3}}^\prime\frac{1}{v_{lnm}}g^{\prime\prime}
(v_{lnm})
\nonumber\\
&-& \frac{\pi}{4 \hbar c} T ^2{\sum_n}^\prime \big[a_1
g^\prime(v_{n00}) +a_2 g^\prime(v_{0n0}) +a_3
g^\prime(v_{00n})\big]~.
\end{eqnarray}
Notice that $\Delta E(T)$  does not depend on the value of $K$. In
fact, this expression can be obtained directly from the spectral
density without any explicit calculation of $\Delta S$ and $\Delta
F$. The regular behavior of $\Delta E$ is the reason why previous
authors have restricted their calculations to this particular
thermodynamical variable \cite{case}.

The behavior of the total entropy and the difference between its
``free space" and bounded space values, is shown in Fig.~2,
together with the free and internal energies. Three particular
configurations were considered for the numerical calculations: the
cases of a  ``pizza box", $a_2=a_3=100a_1$; a cube, $a_1=a_2=a_3$;
and a wave-guide configuration, $a_2 =a_3=a_1/10$. Some common
features are worth noticing: (i) the total entropy $S$ and the
relative entropy $\Delta S$ are zero at $T=0$, as it should be by
construction; (ii)the total entropy is always positive; (iii)the
relative entropy may take negative values and, accordingly, heat
may be released by the system when going from free space to a
bounded configuration; (iv) the entropy and free energy of the
system are discontinuous functions of the temperature; (v) the
free and internal energies equal the Casimir energy at $T=0$; (vi)
all relative thermodynamic variables tend to zero as
$T\rightarrow\infty$.

The case $a_2=a_3=100a_1$ is similar to the widely studied
parallel plates configuration; it must be realized, however, that
there is a qualitative difference between the two configurations:
in closed rectangular boxes, all frequencies are discrete and {\bf
no} zero frequency electromagnetic modes are allowed, whereas
parallel plates admits a continuous range of frequencies and the
existence of either $TE$ and $TM$ of $zero$ frequency is not
excluded in principle\cite{milton}. The presence of such modes
significantly alters the behavior of the free and internal
energies at high temperatures and introduce terms that increase
linearly with temperature \cite{brown,boyer,milton}. Our
calculations show that for a closed rectangular cavity, such
linear terms are absent and that, moreover, all the effects
associated to the discrete density of modes $\Delta\rho(\omega)$
tend to disappear as the temperature increases. The case of a
cube, $a_1=a_2=a_3$, is particularly interesting since its Casimir
energy is positive; in this case, the difference between the free
and bounded internal energy $\Delta E$ is a decreasing function of
the temperature. In the case $a_1=a_2=a_3/10$, the cavity is
similar to a waveguide and discontinuities of $F$ and $S$ persist
at higher temperatures.

Once a regularized expression for the thermodynamic potentials has
been obtained, other relevant physical variables can be directly
calculated. For instance, the specific heat $C_V$ for a given
geometry,
\begin{equation}
              C_V =\Big(\frac{\partial E}{\partial T}\Big)_V
\end{equation}
is clearly a continuous function of $T$. For a cube, it turns out
that the presence of the boundaries reduces the specific heat
$C_V$ with respect to the case with boundaries at infinity, while
the opposite effect occurs for the ``pizza box", $a_1 \ll
a_2=a_3$. As a general feature, $C_V$ decreases  at very low
temperatures with respect to the free space configuration, but
increases at moderate temperatures.

The pressure on wall $1$ of area $a_2a_3$ is given by
\begin{equation}
P_1 = -\frac{1}{a_2a_3}\Big(\frac{\partial F}{\partial
a_1}\Big)_T~,
\end{equation}
with similar expressions for $P_2$ and $P_3$. A straightforward
calculation shows that, as expected, the equation of state
\begin{equation}
E = (P_1+P_2+P_3)V
\end{equation}
is satisfied. For a cube $P_1=P_2=P_3=E/3V$ and the pressure is a
continuous functions of $T$ . For other configurations, the
regularized expression of $F$ leads to discontinuities of the
pressures at the walls at low temperature. This can be seen in
Fig.~3, where a plot is shown of the total pressure at walls $1$
and $3$ for ``waveguide" and ``pizza box" configurations. It is
worth recalling that the pressure is the physical variable most
accessible to experimental verification.

A common feature of some thermodynamic variables for all
configurations is the appearance of discontinuities as the
temperature varies. These discontinuities are due to the different
weights that the EM  modes acquire as the temperature increases.

\section{Other approaches}

In this section we compare our results with those reported in the
literature for the same or similar systems. For this purpose, it
is convenient to use the formula
\begin{equation}
\coth \pi x -\frac{1}{\pi x} = \frac{2 x}{\pi}\sum_{k=1}^\infty
\frac{1}{x^2 + k^2}
\end{equation}
and write the regularized free energy in the form
\begin{eqnarray}
\Delta F =  \Delta E_0 + \Delta F_M +\frac{V}{2\pi
c^3}T \sum_{|v_{\bf n}| > v_V}\frac{1}{u_{\bf n}^3}\nonumber\\
- \frac{\pi}{4}T\Big[\sum_{v_{n00}>v_E} \frac{1}{n} ~+
\sum_{v_{0n0}>v_E} \frac{1}{n} ~+ \sum_{v_{00n}>v_E}
\frac{1}{n}\Big]~,
\end{eqnarray}
where $\Delta F_M$ is the free energy that follows directly from
the Matsubara formalism, see $e.$ $g.$ Santos and
Tort\cite{santos}. It can be obtained in our own formalism using
$\Delta\rho$ with the integration paths so chosen that $K_V$ and
$K_E$ are taken as zero for every $v_{nlm}$. Explicitly:
\begin{eqnarray}
\Delta F_M &=& -\frac{2V \hbar}{\pi c^3} {\sum_{\bf
n}}^\prime\sum_{k=1}^\infty \frac{1}{[(k\hbar / T)^2 + u_{\bf
n}^2]^2 } +\frac{\hbar a_1 }{c}\sum_{n,k=1}^\infty
\frac{1}{(k\hbar /T )^2 + u_{n00}^2 }\nonumber\\ &+&\frac{\hbar
a_2 }{c}\sum_{n,k=1}^\infty \frac{1}{(k\hbar /T )^2 + u_{0n0}^2
}+\frac{\hbar a_3 }{c}\sum_{n,k=1}^\infty \frac{1}{(k\hbar /T )^2
+ u_{00n}^2 }. \label{eq:matsu}
\end{eqnarray}
Notice that this expression has a logarithmic divergence and,
consequently, it precludes any practical calculation of physical
quantities. The analogue of $\Delta F_M$ for a massless scalar
field was reported in Ref. \cite{ambjorn}, where, as mentioned
above in the Introduction, the problem of the divergence was
stated but no explicit finite expression for the free energy,
valid at all temperatures, was given.

From the explicit form of the terms that depend on $K_{V,E}(v)$ in
Eq. (\ref{eq:deltaF}), it can be seen that they have a quantum
origin: although each term in the summation is independent of
$\hbar$, the selection of the integers $n_i$ that contribute to
the summation depends on $\hbar$. Indeed, Santos and Tort
\cite{santos} have shown that the free energy of the EM field
calculated using the $Z$-function regularization technique is
identical with $\Delta F_M$ up to a term $T {\rm ln}
(\mu/\sqrt{2\pi}~T)$, where $\mu$ is a scale factor. Santos and
Tort choose $\mu = \sqrt{2\pi}~T$, whereas in our regularization
scheme
\begin{equation}
\mu = \sqrt{2\pi}~T \exp\Big\{\frac{V}{2\pi c^3}\sum_{|v_{\bf
n}|>v_V}\frac{1}{u_{\bf n}^3} -
\frac{\pi}{4}\Big[\sum_{v_{n00}>v_E} \frac{1}{n} ~+
\sum_{v_{0n0}>v_E} \frac{1}{n} ~+ \sum_{v_{00n}>v_E}
\frac{1}{n}\Big]\Big\}
\end{equation}
is the scale factor.

Since it is well known  that most difficulties with the
quantization of the electromagnetic field are related to the fact
that it is a massless field, the origin of our divergence problem
can also be clarified by assigning an effective mass $m_\gamma$ to
the photon. For a massive field, the volumetric contribution to
$\Delta F_M$ must be changed to \cite{ambjorn}:
\begin{equation}
\Delta F_M(m_\gamma) = -  \frac{ Vm_{\gamma}^2 c}{4\pi^2 \hbar}
{\sum_{\bf n}}^\prime\sum_{k=1}^\infty  \frac{K_2 \Big(
(m_{\gamma} c^2 / 2\hbar) \Big[(\hbar  k/ T)^2 + u_{\bf n}^2
\Big]^{1/2}\Big)}{(\hbar k/ T)^2 + u_{\bf n}^2 } ~, \label{Fbb2}
\end{equation}
where $K_2$ is the associated Bessel function. The limit of low
temperature, $T \rightarrow 0$, of this expression can be
calculated as a continuous integral over variables $n_i$, with the
result
\begin{equation}
\Delta F_M(m_\gamma) \simeq - T \ln\Big(1 - e^{-m_{\gamma} c^2
/2T}\Big) ~,
\end{equation}
from where the following contribution to the entropy is obtained:
\begin{equation}
\Delta S_M(m_\gamma) = - \frac{\partial F_{BB}}{\partial T} =
\ln\Big(1 - e^{-m_{\gamma} c^2/2T}\Big) + \frac{m_{\gamma} c^2
}{2T}~~\frac{1}{e^{m_{\gamma}c^2 /2T} - 1}~,
\end{equation}
which is finite and does tend to zero in the limit of low
temperature. Clearly this limit is to be understood in the sense
that $T \ll m_{\gamma}c^2$. Had we taken $m_{\gamma}=0$ from the
beginning, the entropy would not be finite; thus the origin of the
divergence can be traced back to the fact that, since the photon
is massless, there is no natural energy for the temperature to be
compared with; this seems to be the origin of the difficulties
with the definition of entropy in closed rectangular cavities. In
principle, one could use Eq. (\ref{Fbb2}) to calculate the
thermodynamic variables and take the limit of zero mass only at
the end, but this procedure is not useful for practical
computations; as far as we know, no expression for the regularized
quantities in closed form  has been obtained in this way.

\section{Summary and conclusions}

In this paper we have obtained fully regularized expressions for
the thermodynamic variables associated to the electromagnetic
field inside a rectangular box with perfect conducting walls. The
regularization procedure we have used is based on the condition
that the entropy be zero at $T=0$, which is equivalent to imposing
the uniqueness of EM vacuum. In this way, we have been able to
calculate thermodynamical quantities that could be tested
experimentally.

In general, infrared divergences in these type of calculations are
due to the possible emission and absorption of an indefinite
number of soft photons. However, in a closed rectangular cavity of
maximum size $L$, the frequencies are discrete and there must be a
lower bound to the frequency, $\omega_{min} = c/2L$, such that the
limit $\omega\rightarrow 0$ is never achieved in the system.
According to our analysis, this seems to be the origin of the
difficulties in the infrared limit. Thus special care must be
taken in counting the different microstates in the evaluation of
the free energy: specifically, it is necessary to impose the
uniqueness of the state without photons of any allowed frequency.
Otherwise, the entropy is not well defined at zero temperature.

In the regularization scheme we propose, we have taken the above
facts into account using a cut-off procedure. According to quantum
statistics, the cut-off values $v_V$ and $v_E$ can be interpreted
as parameters that define which EM modes are compatible with the
macrostate of a given configuration, that is, those that must be
properly counted at low temperatures. In particular, it turns out
that the cut-off  related to the volume and the edges of the
configuration are different: $v_V \neq v_E$.

The introduction of cut-off terms is also related to the fact that
all energy fluctuations induced by the thermal bath must be larger
than the quantum fluctuations. Actually, this is the meaning of
the important inequality (\ref{desig}). On the other hand, if the
photons had an effective mass $m_\gamma \sim \hbar /Lc$, no
cut-off would be necessary.

The discontinuities exhibited by the free energy and the entropy,
associated with the modes that fit into the configuration as the
temperature varies, are some of the most striking predictions of
our calculations. However, they could be due to an excessive
idealization of the system under study, namely, a perfectly
conducting cavity at thermal equilibrium with its surroundings. It
may happen that for a dielectric closed box such discontinuities
are softened, but traces of them  could be present in experiments.

\begin{acknowledgments}
 This work was partially supported by
Conacyt, M\'exico, under grant 41048-A1 and DGAPA IN-18605-3   .
\end{acknowledgments}

\section*{Appendix}

A basic integral to be evaluated is:
\begin{eqnarray}
&&-\frac{u}{\beta \hbar} \int _{0}^\infty d\omega \cos(u
\omega)\ln \Big(1-e^{-\beta \hbar \omega}\Big)\nonumber\\
&=& \int _{0}^\infty d\omega \frac{\sin(u \omega)}{e^{\beta
\hbar \omega}-1} \nonumber\\
&=& \frac{1}{4}\int_{-\infty}^\infty d\omega \sin(u \omega)
\coth(\beta \hbar \omega /2)
-\frac{1}{2}\int_0^\infty d\omega \sin (u \omega)\nonumber\\
&=&\frac{\pi}{2\beta \hbar}\Big[\coth \Big(\frac{\pi u}{\beta
\hbar}\Big) - K\Big]-\frac{1}{2} {\cal
P}\frac{1}{u}~\label{eq:path}
\end{eqnarray}
(for $u \ne 0$). The value of this integral depends on the
integration path used to circumvent the pole at $\omega=0$. For a
given $u
>0$, one finds that $K= \pm 1$ if the path circumvents the
singularity over (under) the real axis, and $K=0$ if the principal
value of the integral is taken.

The other basic integral is
$$
-\frac{u}{\beta \hbar} \int _{0}^\infty d\omega \omega \sin(u
\omega)\ln \Big(1-e^{-\beta \hbar \omega}\Big)
$$
\begin{equation}
= -\int _{0}^\infty d\omega \omega \frac{ \cos(u \omega)}{e^{\beta
\hbar \omega}-1}+ \frac{1}{u}\int _{0}^\infty d\omega \frac{\sin(u
\omega)}{e^{\beta \hbar \omega}-1}~,
\end{equation}
with
\begin{eqnarray}
\int _{0}^\infty d\omega ~\omega \frac{\cos(u \omega)}{e^{\beta
\hbar \omega}-1} &=& \frac{1}{4}\int_{-\infty}^\infty d\omega
~\omega \cos(u
\omega) \coth(\beta \hbar \omega /2)-\nonumber\\
&&\frac{1}{2}\int_0^\infty d\omega ~\omega \cos(u \omega)\nonumber\\
&=&-\frac{1}{2} \Big(\frac{\pi}{\beta \hbar}\Big)^2 {\rm cosech}^2
\Big(\frac{\pi u}{\beta \hbar}\Big) + \frac{1}{2} {\cal
P}\frac{1}{u^2}~.\label{eq:path2}
\end{eqnarray}
Notice that the value of this last integral is independent of the
integration path around the origin, $\omega = 0$, because the
residue at that point is zero.

\newpage
\begin{center}
{\Large \bf Figure Captions}
\end{center}

\vspace{0.5cm}

{\bf Figure 1.} Plot of the monotonic increasing function
$G(v_0)$. Numerical analysis shows that $G(v_V)=0$ for
$v_V=1.763876988$ and $G(v_E)=-1$ for $v_E=0.64889408$ (with ten
significant figures).

\vspace{0.5cm}

{\bf Figure 2.} Dimensionless thermodynamic potentials $S/k_B$,
$\Delta S/k_B$, $\Delta f=\pi a_1 \Delta F/\hbar c$, and $\Delta
u=\pi a_1 \Delta E/\hbar c$ as functions of the dimensionless
variable $\xi = \pi k_B T a_1/\hbar c$. Three particular
configurations are shown:  $a_2=a_3=100a_1$ (pizza box),
 $a_1=a_2=a_3$ (cube),and $a_2 =a_3=a_1/10$ (wave
guide), in the first, second and third columns respectively. The
total entropy is given in the first row; notice that it is always
positive, which is not the case for the difference between the
entropy in ``free space" and its finite domain value, as seen in
the second row . The finite domain free and internal energies are
shown in the third and fourth rows respectively. In all cases, the
finite domain thermodynamic potentials have the physically
expected values as $\pi k_B T a_1/\hbar c \rightarrow 0$, and tend
to zero as $\pi k_B T a_1/\hbar c \rightarrow \infty$.

\vspace{0.5cm}

{\bf Figure 3.} Dimensionless pressures on the walls 1 and 3,
$p_1=\pi a_1^4 P_1/\hbar c$ and $p_3=\pi a_1^4 P_3/\hbar c$ as
function of the dimensionless variable $\xi = \pi k_B T a_1/\hbar
c$ for a
 ``pizza box" configuration (upper panel), and a ``waveguide" configuration (lower panel).
For comparison the pressures corresponding to $F_{BB}$ are plotted
in dotted lines.

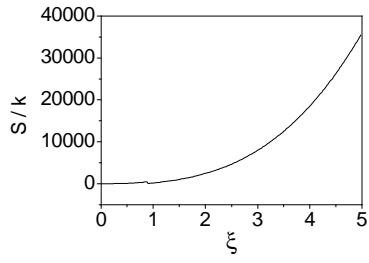
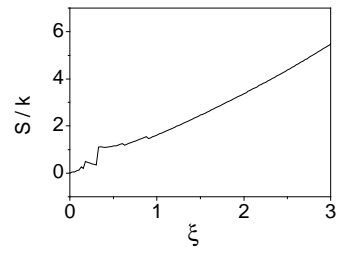
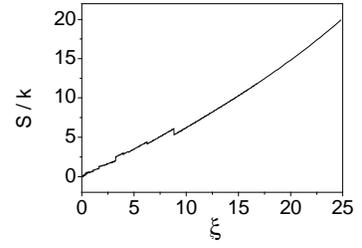
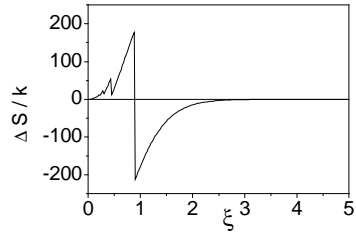
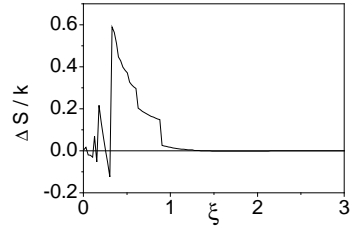
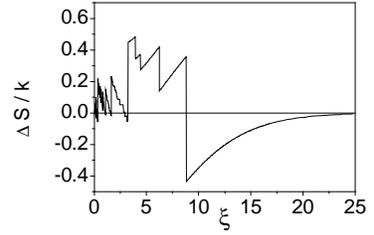
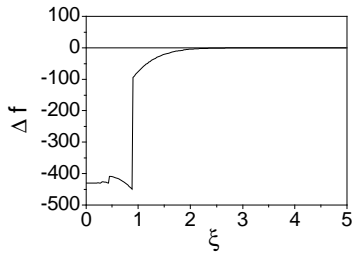
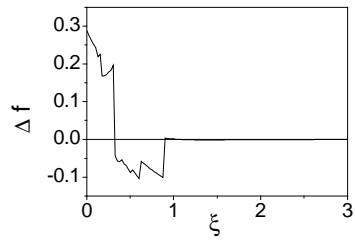
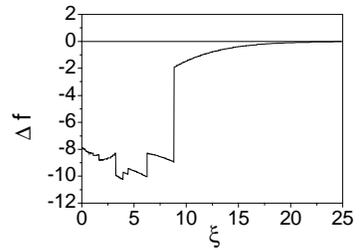
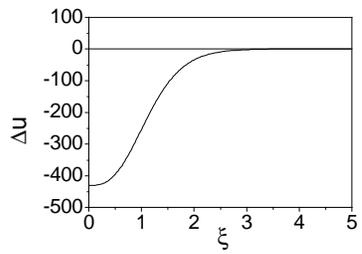
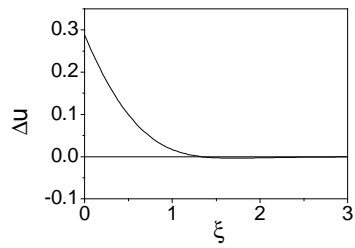
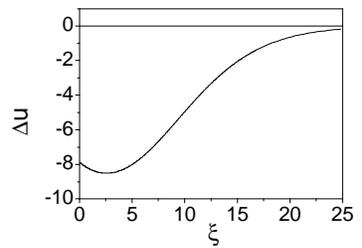
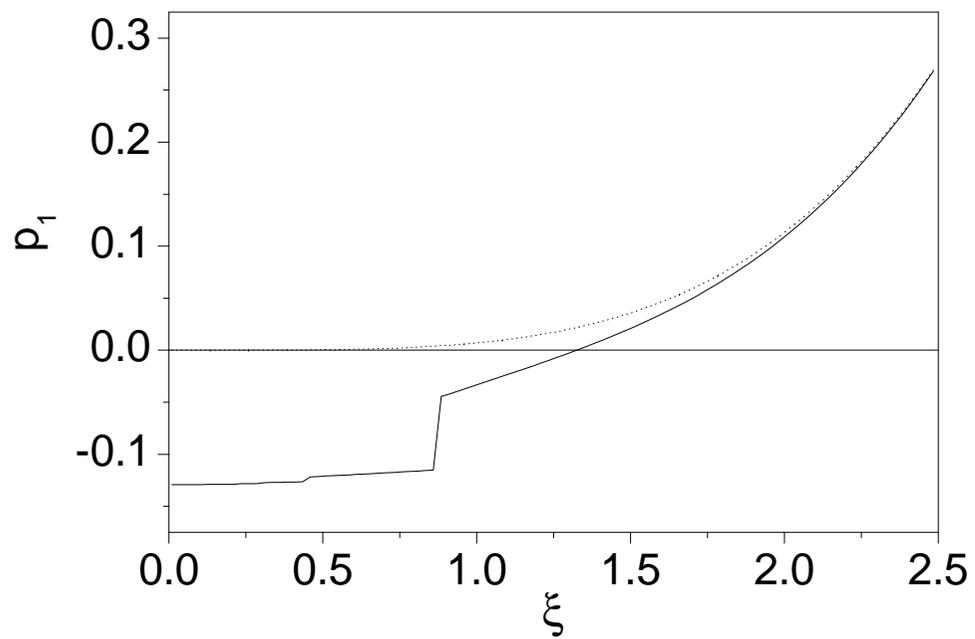
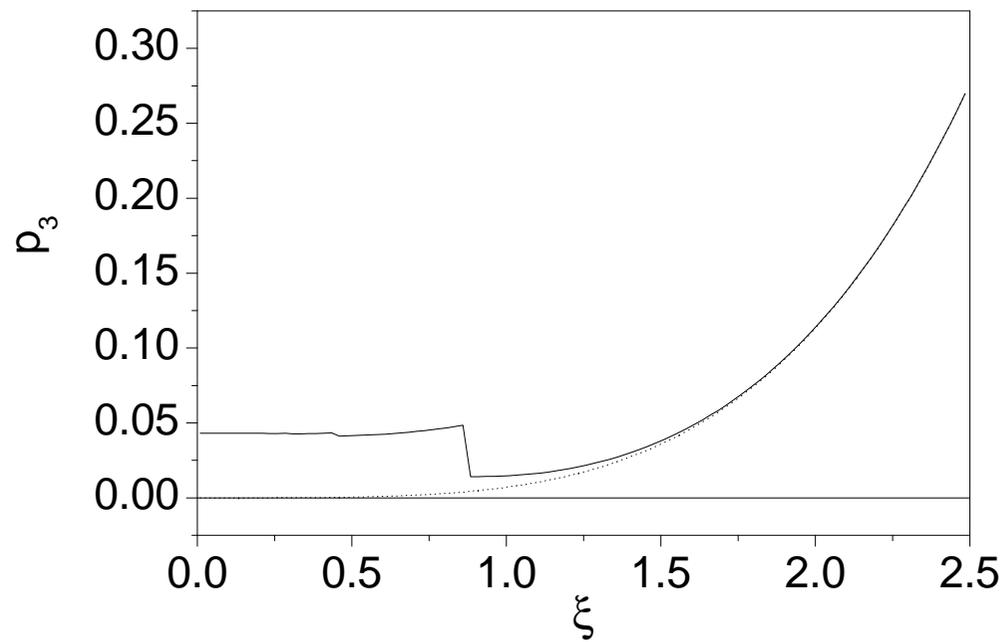
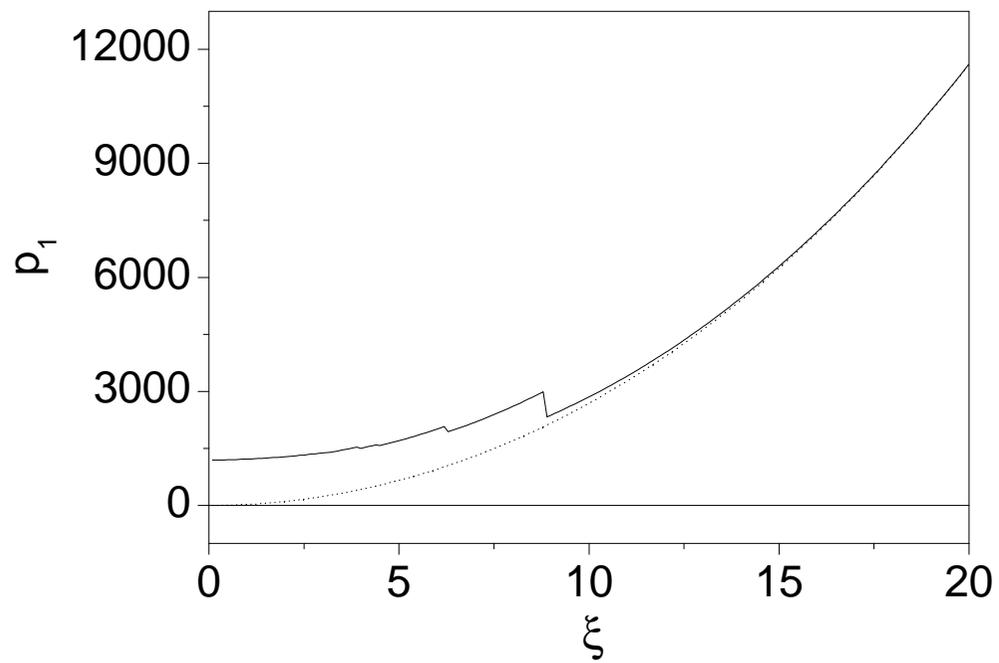
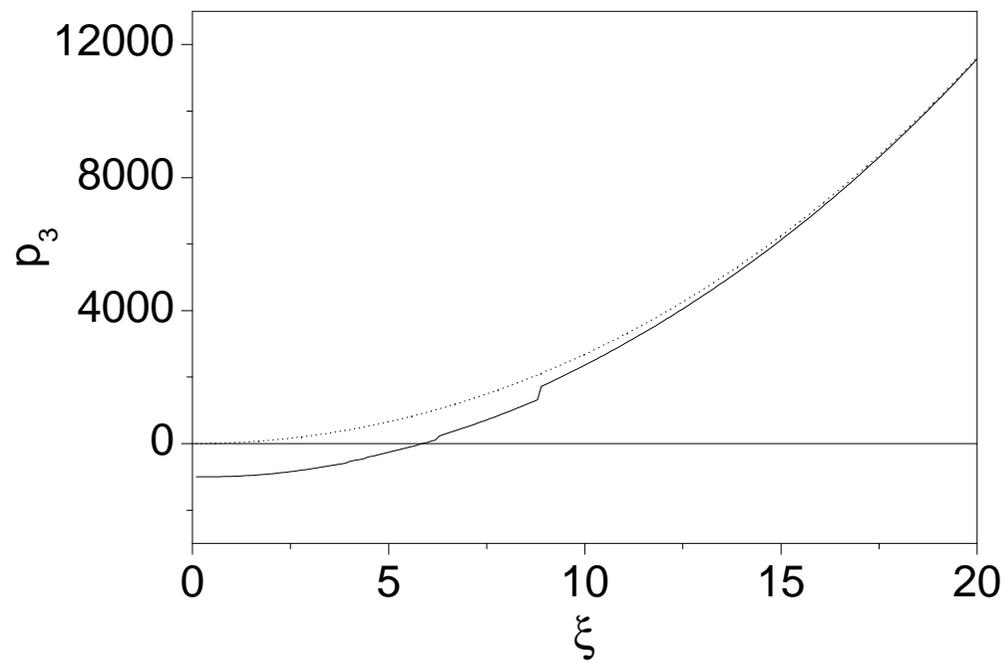
\end{document}